\documentclass[aps,prl,twocolumn,superscriptaddress,footinbib,longbibliography,amsmath,amssymb]{revtex4-2} 

\usepackage{graphicx}
\usepackage{subfigure}
\usepackage{epsfig} 
\usepackage{dcolumn}
\usepackage{bm}
\usepackage{times} 
\usepackage{color}

\begin{document}

\title{Twist-induced control of near-field heat radiation between magnetic Weyl semimetals}

\author{Gaomin Tang}
\email{gaomin.tang@unibas.ch}
\affiliation{Department of Physics, University of Basel, Klingelbergstrasse 82, CH-4056
Basel, Switzerland}
\author{Jun Chen}
\affiliation{State Key Laboratory of Quantum Optics and Quantum Optics Devices, Institute 
of Theoretical Physics, Shanxi University, Taiyuan 030006, China}
\author{Lei Zhang}
\email{zhanglei@sxu.edu.cn}
\affiliation{State Key Laboratory of Quantum Optics and Quantum Optics Devices, Institute
of Laser Spectroscopy, Shanxi University, Taiyuan 030006, China}
\affiliation{Collaborative Innovation Center of Extreme Optics, Shanxi University, Taiyuan
030006, China}

\bigskip

\begin{abstract}
  Due to the large anomalous Hall effect, magnetic Weyl semimetals can support
  nonreciprocal surface plasmon polariton modes in the absence of an external magnetic
  field. This implies that magnetic Weyl semimetals can find novel application in
  (thermal) photonics. In this work, we consider the near-field radiative heat transfer
  between two magnetic Weyl semimetal slabs and show that the heat transfer can be
  controlled with a relative rotation of the parallel slabs. 
  Thanks to the intrinsic nonreciprocity of the surface modes, this so-called twisting
  method does not require surface structuring like periodic gratings. The twist-induced
  control of heat transfer is due to the mismatch of the surface modes from the two slabs
  with a relative rotation.
\end{abstract}

\maketitle

{\it Introduction.--}
Near-field radiative heat transfer (NFRHT) can largely exceed the Planckian limit of
black-body radiation~\cite{Planck} due to the contribution from surface
electromagnetic modes~\cite{near1, PvH, Loomis94, Xu94, review05, review07, review15,
review15-2, review18, GT1, Zhang_book, review20} and attracts particular scientific
interest triggered by experimental advances~\cite{Gelais14, Song15, Kim15, Song16,
Bernardi16, Cui17, Kloppstech17, Ghashami18, Fiorino18, DeSutter19, Tang20}. 
For novel applications, it is of importance to actively control NFRHT.
Several strategies have been proposed, such as applying an electric
field to phase-change materials~\cite{control_electric_11} or ferroelectric
materials~\cite{control_electric_14}, applying an external magnetic field to
magneto-optical materials~\cite{control_magnetic_15, control_magnetic_17,
control_magnetic_18, control_magnetic_19, control_magnetic_20, control_magnetic_21},
drift currents~\cite{drift1, drift2}, and regulating the chemical potential of
photons~\cite{control_potential_15}. 
Another active control strategy is to utilize the rotational degree of
freedom~\cite{twist_radiate2011, twist_radiate2017, SPP-BP, twist_radiate1,
twist_radiate2, twist_radiate3, twist_radiate4, GT_SMP}. 
In analogy to the twistronic concept in low-dimensional materials~\cite{twist_Cao_1,
twist_Cao_2, Chen983} and photonics~\cite{twist_Ye_1, twist_Ye_2, twist_Alu_1,
twist_Alu_2}, this control strategy is also called twisting method.
So far, most of the proposals for the realizations of the 
twisting method require nanometer-sized periodic
gratings to create anisotropic patterns~\cite{twist_radiate2011, twist_radiate1,
twist_radiate2, twist_radiate3, twist_radiate4}. 

Due to inherent time-reversal symmetry breaking, magnetic Weyl semimetals (WSMs), such as
Co$_3$Sn$_2$S$_2$~\cite{WSM_AHE_18nc, WSM_AHE_18np}, Ti$_2$MnAl~\cite{WSM_AHE_18PRB},
EuCd$_2$As$_2$~\cite{WSM_AHE_19PRB}, Co$_2$MnGa~\cite{WSM_room19}, and
Co$_2$MnAl~\cite{WSM_AHE_20nc}, can exhibit
large anomalous Hall effect so that the dielectric tensor has large off-diagonal
components. This leads to the existence of nonreciprocal surface plasmon polaritons
(SPPs)~\cite{WSM_SPP16, Kotov18, WSM_SPP19_2, WSM_SPP19_3} and breaks the Lorentz
reciprocity. 
The broken Lorentz reciprocity violates Kirchhoff's law of radiation and opens
opportunities for a variety of radiative applications~\cite{nonreciprocal_14,
nonreciprocal_PNAS, nonreciprocal_18, nonreciprocal_19, Khandekar19, nonreciprocal_20,
nonreciprocal_diode}. 
Compared to magneto-optical materials, magnetic WSMs break Lorentz reciprocity
intrinsically in the absence of external magnetic fields and this has been studied from the
perspective of (thermal) radiation very recently~\cite{WSM_radiate1, WSM_radiate2,
WSM_radiate3, WSM_radiate4, WSM_radiate5}. 
Moreover, it has been shown that magnetic WSMs can exhibit nonreciprocal reflectivity
without surface structuring using a planar interface~\cite{WSM_radiate3}. 

In this Letter, we employ the intrinsic nonreciprocity of the surface modes in
magnetic WSMs and demonstrate that NFRHT between magnetic WSMs can be actively controlled
via twist. 
We will first show how the nonreciprocal dispersion of SPPs changes with the incidence
plane of the light. Using fluctuational electrodynamics, we will study the implications of
nonreciprocity on NFRHT and the twisting effects between two WSM slabs. 

{\it Surface plasmon polaritons.--}
In WSM, either inversion or time-reversal symmetry needs to be broken to split a doubly
degenerate Dirac point into a pair of Weyl nodes with opposite chirality~\cite{WSM_Wan11,
WSM_RMP}.  
Each pair of Weyl nodes are separated in momentum space (denoted by wave vector $2{\bf
b}$) by breaking time-reversal symmetry or with an energy of $2\hbar b_0$ by breaking
inversion symmetry. 
The presence of Weyl nodes changes the electromagnetic response and the displacement
electric field for WSM in the frequency domain is written as~\cite{WSM_chiral_12}
\begin{equation} \label{D}
  {\bf D} = \epsilon_0\epsilon_d{\bf E} +\frac{ie^2}{4\pi^2\hbar\omega}(-2b_0{\bf B}+2{\bf
  b}\times{\bf E})
\end{equation}
with $\omega$ the angular frequency.
The dielectric function $\epsilon_d$ is expressed as 
$\epsilon_d =\epsilon_b +i\sigma/\omega$
where $\epsilon_b$ is the background permittivity and $\sigma$ the bulk conductivity. 
It is seen from Eq.~\eqref{D} that $b_0$ gives rise to the chiral magnetic effect and
${\bf b}$ the anomalous Hall effect. 
This implies that magnetic WSMs with broken time-reversal symmetry can give rise to
the anomalous Hall effect.  
Considering $2{\bf b}$ along the $y$-direction in momentum space (${\bf b}=b\hat{q}_y$)
and inversion symmetric system with $b_0=0$, 
we have ${\bf D} =\epsilon_0\bar{\bar{\epsilon}}{\bf E}$ in the Cartesian coordinate
system where the dielectric tensor is
\begin{equation}
  \bar{\bar{\epsilon}}(\omega) = 
  \begin{bmatrix}
    \epsilon_d & 0 & i\epsilon_a \\
    0 & \epsilon_d & 0 \\
    -i\epsilon_a & 0 & \epsilon_d 
  \end{bmatrix} 
\end{equation}
with $\epsilon_a = be^2/(2\pi^2\epsilon_0\hbar\omega)$. 
It has been reported that $\epsilon_a$ can be comparable to $\epsilon_d$ in the infrared
region which is of most interest for thermal applications~\cite{WSM_radiate1,
WSM_radiate2, WSM_radiate3, WSM_radiate4}. 

\begin{figure}
\centering
\includegraphics[width=\columnwidth]{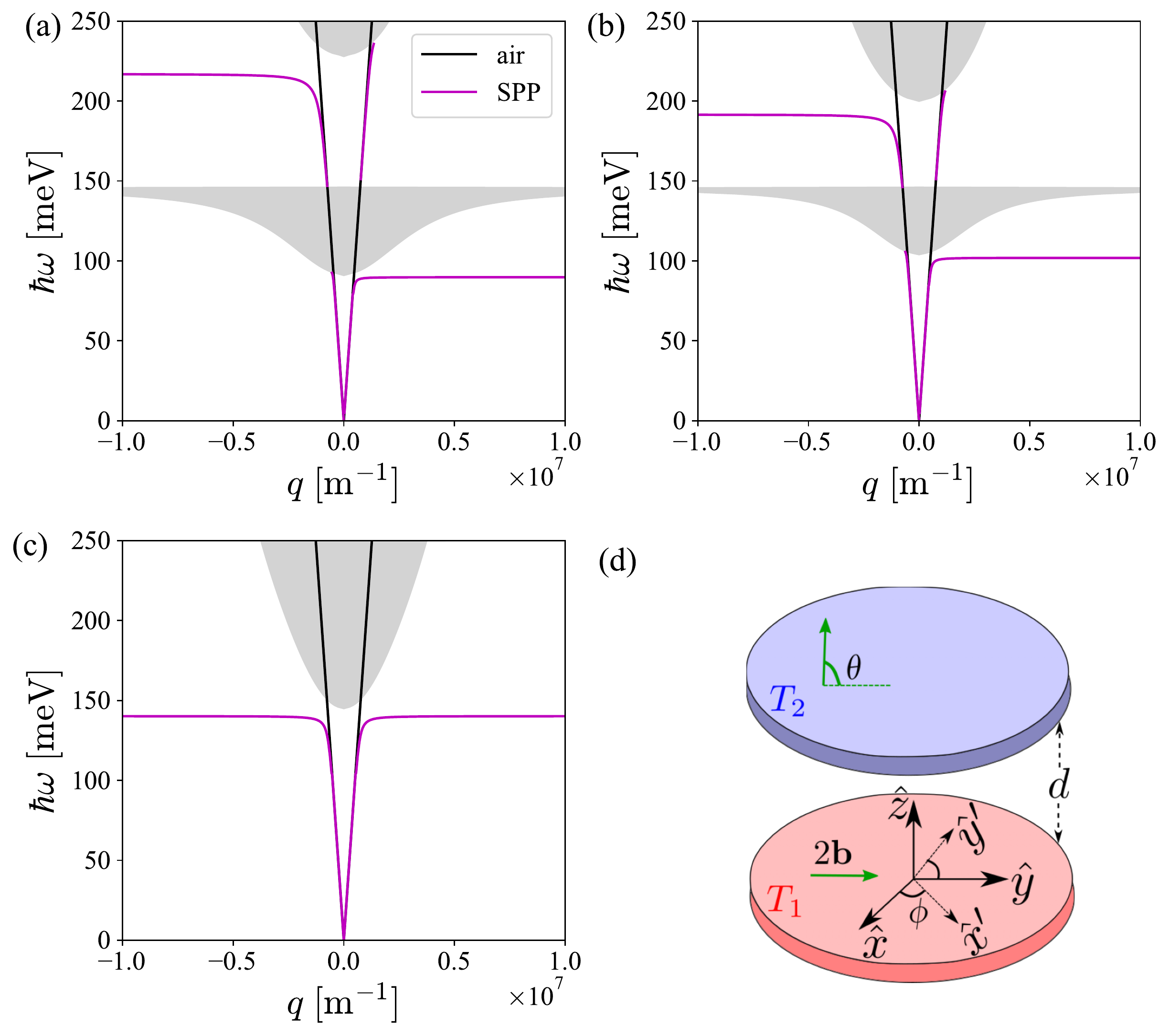} \\
\caption{The dispersion of surface plasmon polaritons (magenta lines) with different
  azimuthal angles of incidence: (a) $\phi=0$, (b) $\phi=\pi/4$, and (c) $\phi=\pi/2$. 
  The black lines are the linear dispersion relation in air (or vacuum). 
  The gray regions show the continua of the bulk plasmon modes in Weyl semimetal. 
  (d) Schematic setup for near-field heat radiation between two Weyl semimetals with gap
  separation $d$ and twist angle $\theta$. The twist angle is defined as the angle between
  the Weyl node separations in the bottom and top Weyl semimetals. }
\label{fig1}
\end{figure}

We first discuss the dispersion relations of SPPs at the planar interface between WSM and
air by considering only one WSM slab. 
With the incidence plane at azimuthal angle $\phi$ with respect to the $x$-axis, which is
the $x'$-$z$ plane shown in Figure~\ref{fig1}(d), the dielectric tensor is transformed to
\begin{equation}
  \bar{\bar{\epsilon}}\,{}'(\omega) =
  {\cal R} \bar{\bar{\epsilon}}(\omega){\cal R}^T
   = 
  \begin{bmatrix}
    \epsilon_d & 0 & i\epsilon_a \cos\phi \\
    0 & \epsilon_d & i\epsilon_a \sin\phi \\
    -i\epsilon_a \cos\phi & -i\epsilon_a \sin\phi & \epsilon_d 
  \end{bmatrix} 
\end{equation}
where ${\cal R}$ is the rotation matrix of angle $\phi$. 
We start from Maxwell curl equations
\begin{equation}
  \nabla \times {\bf E} = -\partial_t {\bf B}, \qquad
  \nabla \times {\bf H} = \partial_t {\bf D},  \label{Maxwell}
\end{equation}
with ${\bf B}=\mu_0\mu{\bf H}$ and ${\bf D}=\epsilon_0\bar{\bar{\epsilon}}\,{}' {\bf E}$. 
Since the SPP is transverse magnetic (or $p$-polarized) mode, the magnetic fields in
air (${\bf H}_0$) and in WSM (${\bf H}_1$) are written in the forms as
\begin{align}
  {\bf H}_0(x',z,t) &=\hat{y}' H e^{iqx' - i\beta_0 z}e^{-i\omega t}, 
  \quad &&{\rm Im}(\beta_0)<0, \\
  {\bf H}_1(x',z,t) &=\hat{y}' H e^{iqx' + i\beta_1 z}e^{-i\omega t}, 
  \quad &&{\rm Im}(\beta_1)<0,
\end{align}
where $q$ is the in-plane wave vector. The out-of-plane wave vectors in air and WSM are
denoted as $\beta_0$ and $\beta_1$, respectively. 
Using Maxwell equations in the WSM and air, respectively, one has
\begin{equation}  
  \beta_0^2 + q^2 = k_0^2, \qquad
  \beta_1^2 + q^2 = \mu \epsilon_{\rm eff} k_0^2, \label{beta}
\end{equation}
with $k_0=\omega/c$ the wave vector in air and the dielectric function
$\epsilon_{\rm eff} = \epsilon_d-(\cos\phi\,\epsilon_a)^2/\epsilon_d$. 
Using the interface condition of electric field, the
implicit dispersion relation for the SPP is obtained as
\begin{equation} \label{dispersion}
  \epsilon_{\rm eff} \beta_0 + \beta_1 +i\cos\phi\, \epsilon_a q /\epsilon_d =0 .
\end{equation}
It can be seen from Eq.~\eqref{dispersion} that the dispersion is nonreciprocal as long as
$\cos\phi\neq 0$ and is reciprocal in the Faraday configurations with $\phi=\pi/2$ or
$\phi=3\pi/2$. 
From Eqs.~\eqref{beta} and \eqref{dispersion}, the dispersion relation of SPP can be
numerically obtained. 
The bulk plasmon dispersion is found as $q = \pm \sqrt{\mu \epsilon_{\rm eff}} k_0$ with
$\epsilon_{\rm eff} >0$. 
We consider the case with the relative permeability $\mu$ to be $1$. 

The bulk conductivity $\sigma$ can be obtained using the Kubo-Greenwood formalism to a
two-band model with spin degeneracy as~\cite{Kotov16,Kotov18}
\begin{align}
  \sigma =\frac{g r_s}{6}\Omega G\bigg(\frac{\hbar\Omega}{2}\bigg) +i\frac{g r_s}{6\pi} 
  \bigg\{ \frac{4}{\hbar^2\Omega} \Big[E_F^2 + \frac{\pi^2}{3} (k_BT)^2 \Big] \notag \\
  + 8\Omega \int_0^{E_c} \frac{G(E)-G(\hbar\Omega/2)}{(\hbar\Omega)^2-4E^2} E dE
  \bigg\} .
\end{align}
Here, $g$ is the number of Weyl nodes, 
$r_s = e^2/(4\pi\epsilon_0\hbar v_F)$ is the effective fine-structure constant with Fermi
velocity $v_F$, 
$\Omega=\omega +i2\pi\tau^{-1}$ with the Drude damping rate $\tau^{-1}$, 
$G(E)=n(-E)-n(E)$ with  the Fermi distribution function $n(E)$,
$E_F$ is the chemical potential, 
and $E_c$ is the cut-off energy. 
Following Refs.~\cite{WSM_radiate1,WSM_radiate2,WSM_radiate3, WSM_radiate4}, we take the
parameters $b=2\times 10^9\,{\rm m}^{-1}$, $\epsilon_b=6.2$, $g=2$, $v_F = 0.83\times
10^5\,$m/s, $\tau=1000\,$fs, $E_F =0.15\,$eV at temperature $T=300\,$K, and $E_c=3E_F$. 
The parameters are close to the reported values for Co$_3$Sn$_2$S$_2$~\cite{WSM_AHE_18nc,
WSM_AHE_18np} and the room temperature WSM Co$_2$MnGa~\cite{WSM_room19}.

Figures.~\ref{fig1}(a)-\ref{fig1}(c) show the dispersions of SPPs at different incidence
planes characterized by the azimuthal angle $\phi$. The gray regions show the continua of
bulk plasmon modes which are reciprocal. 
At $\phi=0$ (Voigt configuration), the nonreciprocity of the SPPs is clearly identified by
the asymmetry with respect to the wave vector $q$. 
There are two continua of bulk plasmon modes: one is lower in frequency and the other
higher. The low-frequency continuum separates the SPPs into two branches. 
With increasing the azimuthal angle from $\phi=0$ to $\phi=\pi/2$ (Faraday configuration),
the low-frequency continuum shrinks and the degrees of nonreciprocity decreases. 
At $\phi=\pi/2$, the low-frequency continuum vanishes and the SPP dispersion becomes
strictly reciprocal. 

Here, we only show the dispersions of SPPs which are $p$-polarized. There exist
$s$-polarized surface modes as well. As it was shown in Ref.~\cite{WSM_radiate4}, the
$s$-polarized modes are nonreciprocal between Voigt and Faraday configurations. 
Since it is the $p$-polarized modes that dominate the NFRHT in WSM, the twist-induced
near-field thermal control is mainly due to the nonreciprocity of SPPs. 

\begin{figure*}
\centering
\includegraphics[width=5.4in]{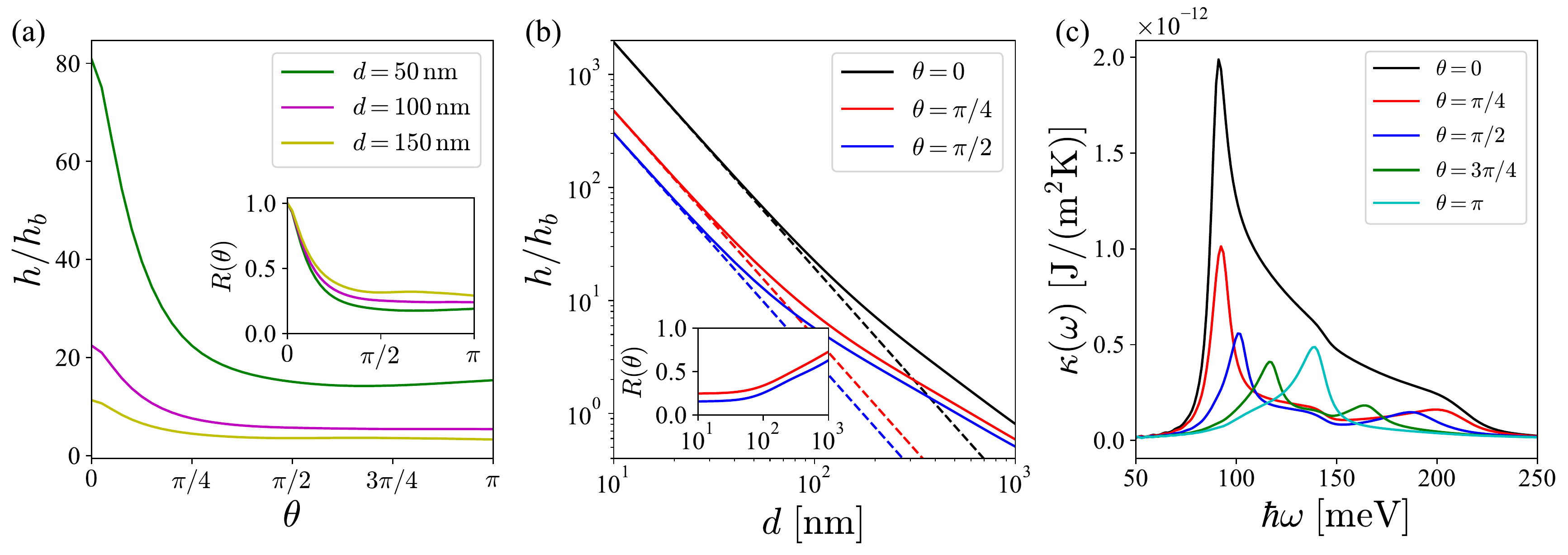} \\
\caption{(a) Scaled heat transfer coefficient $h/h_b$ versus twist angle $\theta$ at
  different gap separations $d$.
  (b) Scaled heat transfer coefficient $h/h_b$ versus gap separation $d$ under different
  twist angles. The corresponding dashed lines are plotted using $h \propto d^{-2}$. The
  thermal switch ratios $R(\theta)$ are shown as inset in (a) and (b).
  (c) Spectral function $\kappa(\omega)$ at different twist angles $\theta$ with
  $d=100\,$nm.}
\label{fig2}
\end{figure*}

{\it Near-field radiative heat transfer.--}
We now consider the NFRHT between two magnetic WSM slabs of the same properties with
temperatures $T_{1(2)}=T\pm \Delta T/2$. The two slabs are placed in parallel and
separated by an air gap with distance $d$ [See Figure~\ref{fig1}(d)]. The twist angle
$\theta$ is the angle between the Weyl node separations in the two slabs and can be
changed by rotating one of the WSMs.  
From the fluctuational electrodynamics~\cite{PvH,Zhang_book}, the radiative heat transfer
coefficient (HTC) $h(\theta)$ at temperature $T$ is given by
\begin{align}
  h(\theta) =\int_0^{\infty} \frac{d\omega}{2\pi} \hbar\omega N' \int_{0}^{\infty}
  \frac{dq}{2\pi} q \int_{0}^{2\pi} \frac{d\phi}{2\pi} \xi(\omega,q,\phi),
\end{align}
where $q$ is the in-plane wave vector and $N'$ is the derivative of Bose-Einstein
distribution $N=1/[e^{\hbar\omega/(k_BT)} -1]$ with respect to the temperature and is
expressed as
\begin{equation}
  N' \equiv \partial N/\partial T =\frac{\hbar\omega\, e^{\hbar\omega/(k_BT)}}{k_B T^2
  \left[ e^{\hbar\omega/(k_BT)} -1 \right]^2}.
\end{equation} 
The photonic transmission coefficient $\xi(\omega,q,\phi)$ is expressed as
\begin{equation}
  \xi = 
  \begin{cases}
    {\rm Tr}[({\bf I}-{\bf R}_2^{\dag}{\bf R}_2){\bf D}({\bf I}-{\bf R}_1{\bf
    R}_1^{\dag}){\bf D}^{\dag}], & q<k_0 \\
    {\rm Tr}[({\bf R}_2^{\dag}-{\bf R}_2){\bf D}({\bf R}_1-{\bf R}_1^{\dag}){\bf
    D}^{\dag}]e^{-2|\beta_0|d}, & q>k_0
  \end{cases} 
\end{equation}
The identity matrix is denoted as ${\bf I}$. 
The reflection coefficient matrix ${\bf R}_n$ at the interface between air and WSM $n$
with $n=1,2$ has the form
\begin{equation} \label{R}
  {\bf R}_n=
  \begin{bmatrix}
    r^{pp}_n & r^{ps}_n \\ r^{sp}_n & r^{ss}_n
  \end{bmatrix} ,
\end{equation}
and is provided in the Supporting Information. 
Furthermore, ${\bf D}=({\bf I}-{\bf R}_1{\bf R}_2 e^{-2i\beta_0 d})^{-1}$ is the
Fabry-Perot-like denominator matrix.
The near- and far-field regimes are defined by the conditions $q>k_0$ and $q<k_0$,
respectively. 
Here, we consider the situation of $T=300\,$K, which can be achieved using room
temperature WSMs discovered recently, such as Co$_2$MnGa~\cite{WSM_room19} and
Co$_2$MnAl~\cite{WSM_AHE_20nc}.
We consider the HTC to be scaled by the corresponding black-body limit $h_b =4\sigma_{\rm
SB} T^3$ with the Stefan-Boltzmann constant $\sigma_{\rm SB} =\pi^2 k_B^4 /
(60\hbar^3c^2)$. One can calculate that $h_b$ is $6.12\,{\rm W/m^2K}$ under $T=300\,$K. 

In Figure~\ref{fig2}(a), we show the scaled HTC $h(\theta)/h_b$ versus the twist angle
$\theta$ at different gap distances $d$. Since $h(\theta)$ is symmetric with respect to
$\theta=\pi$, only the part of $\theta \in [0,\pi]$ is shown. 
The HTC is maximal at $\theta=0$, decreases with increasing $\theta$ to $\theta=\pi/2$ and
remains almost unchanged for $\pi/2 \le \theta \le \pi$.   
The corresponding thermal switch ratios, which are defined as $R(\theta) = h(\theta)
/ h(\theta=0)$, are shown as an inset. Compared to the HTC, the thermal switch ratio is
less sensitive to the gap distance. 
The tunability reported here can be comparable to those by gratings
~\cite{twist_radiate2011, twist_radiate2, twist_radiate3, twist_radiate4} and by rotating
a magnetic field in the case of magneto-optical materials~\cite{control_magnetic_18}. 
Figure~\ref{fig2}(b) shows the dependence of HTC on gap separation $d$. The heat transfer
diverges as $d^{-2}$ at very small distances as shown in dashed lines which was predicted
by Loomis and Maris~\cite{Loomis94}.

\begin{figure}
\centering
\includegraphics[width=\columnwidth]{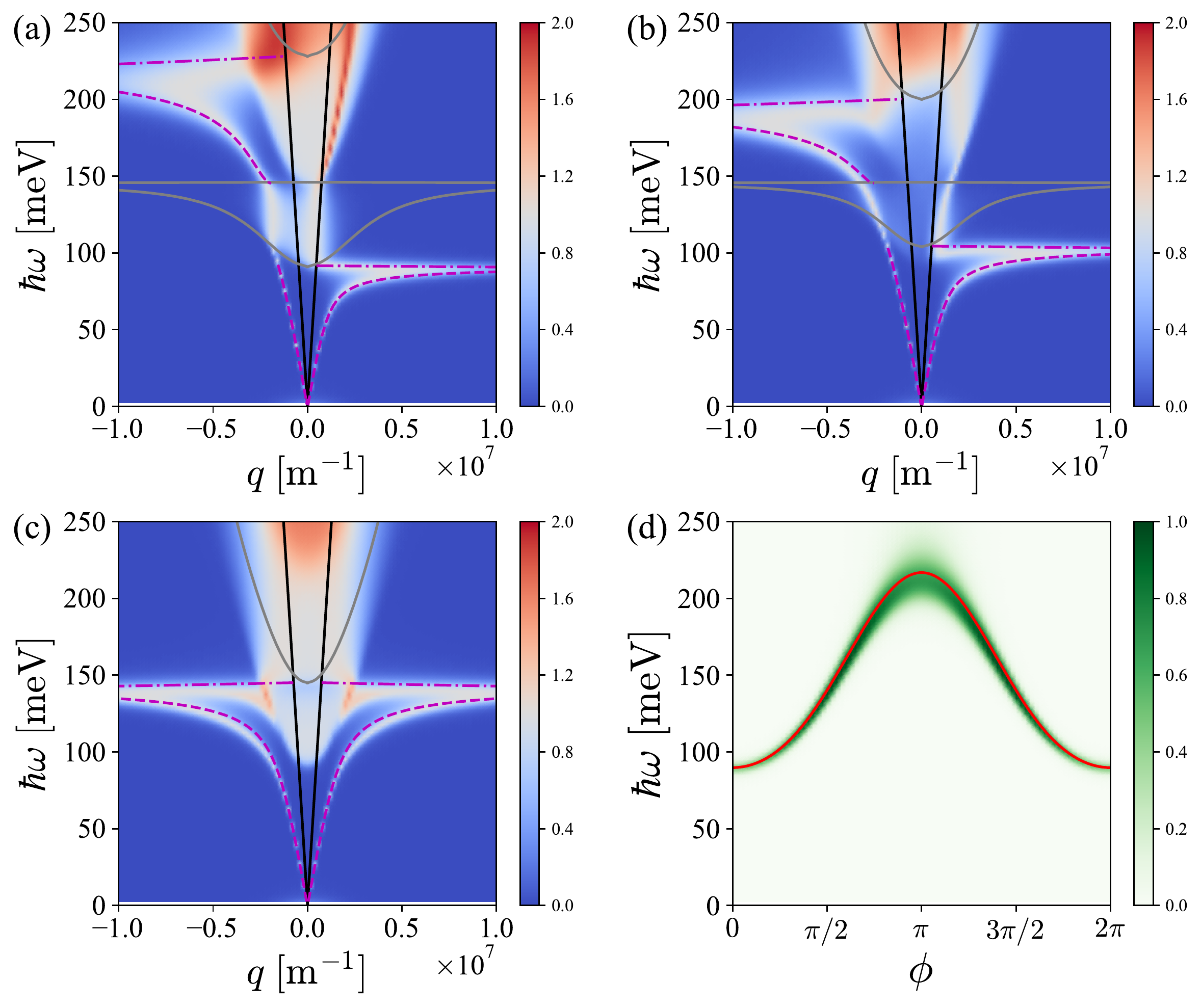} \\
\caption{The photonic transmission coefficients $\xi(\omega,q,\phi)$ are plotted against
  $\hbar\omega$ and $q$ for different azimuthal angles of incidence without twisting
  ($\theta=0$): (a) $\phi=0$, (b) $\phi=\pi/4$, and (c) $\phi=\pi/2$. 
  The black lines depict the linear dispersion in air (or vacuum). The gray lines mark the
  continuum regions of the bulk plasmon modes as in Figures~\ref{fig1}(a)-\ref{fig1}(c). 
  The dashed and dash-dotted magenta lines indicate the odd and even surface plasmon
  polariton modes, respectively.  
  (d) $\xi(\omega,q,\phi)$ plotted against $\hbar\omega$ and $\phi$ at twist angle
  $\theta=0$ and $q=10^7\,{\rm m^{-1}}$. The red line is obtained using
  Eq.~\eqref{dispersion}. The gap separation is $d=100\,$nm.}
\label{fig3}
\end{figure}

The spectral function $\kappa(\omega)$ of HTC is defined through $h =\int_0^{\infty}
\kappa(\omega) d\omega$ and its behaviors for different twist angles $\theta$ are shown in
Figure~\ref{fig2}(c). 
We first focus on the parallel case ($\theta=0$), of which the photonic transmission
coefficients $\xi(\omega,q,\phi)$ against $\hbar\omega$ and $q$ for different $\phi$ in
Figures~\ref{fig3}(a)-\ref{fig3}(c) with $d=100\,$nm. 
Close to or in the far-field regions, $\xi(\omega,q,\phi)$ are less than or equal to $2$,
which is due to the contributions from both $p$- and $s$-polarized modes. 
The contributions to $\xi$ in near field are dominated by the SPPs which are $p$-polarized. 
This is confirmed by Figure S1 in the Supporting Information where the contribution
from the $p$-polarized mode [$r^{pp}_n$ in Eq.~\eqref{R}] on the spectral function is very
close to that from all modes.
For $\theta=0$, the individual SPP from the two WSMs are identical so that
they couple with each other for the whole range of $\phi$ with $\phi \in [0,2\pi]$.
This explains that HTC is maximal at $\theta=0$. 
The near-field regions where $\xi$ are close to $1$ are consistent with the odd (dashed
lines) and even (dash-dotted lines) SPP modes, which are given by
\begin{align} 
  \epsilon_{\rm eff}\beta_0 + \coth(|\beta_0|d/2)(\beta_1 
  +i\cos\phi\,\epsilon_a q /\epsilon_d) =0,\\
  \epsilon_{\rm eff}\beta_0 + \tanh(|\beta_0|d/2)(\beta_1 
  +i\cos\phi\,\epsilon_a q /\epsilon_d) =0,
\end{align}
respectively. Similarly to Figures~\ref{fig1}(a)-\ref{fig1}(c), the degrees of
nonreciprocity for both the odd and even modes decrease from $\phi=0$ to $\phi=\pi/2$ at
which the modes become reciprocal. 
Due to the nonreciprocity of the SPPs, the resonant frequency ranges are different for
different $\phi$ with $0\le\phi\le\pi$ at a given wave vector $q$. This can be seen from
Figure~\ref{fig3}(d) where $\xi(\omega,q,\phi)$ is shown against $\hbar\omega$ and $\phi$ at
$q=10^7\,{\rm m^{-1}}$ under $d=100\,$nm. 
Because of the large nonreciprocity of the SPPs, the whole resonant frequency range is
very broad (from about $90\,$meV at $\phi=0$ to about $220\,$meV at $\phi=\pi$). This can
be seen from the spectral function at $\theta=0$ shown in Figure~\ref{fig2}(c) as well.

\begin{figure}
\centering
\includegraphics[width=\columnwidth]{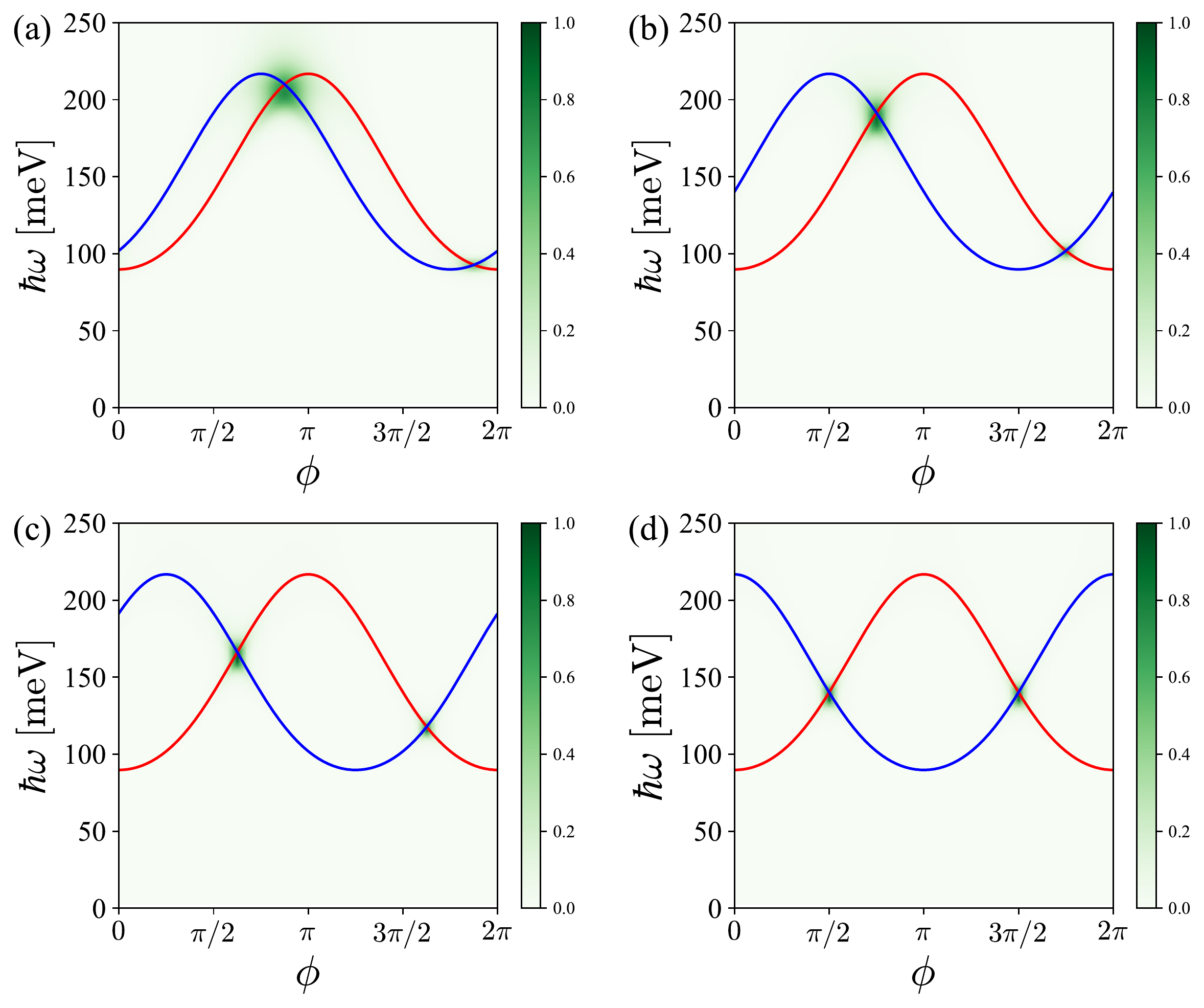} \\
\caption{The photonic transmission coefficients $\xi(\omega,q,\phi)$ are plotted against
  $\hbar\omega$ and $\phi$ at different twist angles with (a) $\theta=\pi/4$, 
  (b) $\theta=\pi/2$, (c) $\theta=3\pi/4$, and (d) $\theta=\pi$ 
  at $q=1/d=10^7\,{\rm m^{-1}}$ with $d=100\,$nm.
  The red lines are the same as the one in Figure~\ref{fig3}(d). The blue lines are obtained
  by shifting the red lines using $\phi \rightarrow \phi+\theta$. }
\label{fig4}
\end{figure}

Now we analyze the twisting effects. 
In Figure~\ref{fig4}, the photonic transmission coefficients $\xi(\omega,q,\phi)$ are
plotted against $\hbar\omega$ and $\phi$ for different twist angles at $q=10^7\,{\rm
m^{-1}}$. The red lines are plotted using the SPP dispersion relation,
Eq.~\eqref{dispersion}, and the blues lines are obtained by performing the shift $\phi
\rightarrow \phi+\theta$. Due to the twist, the red and blue lines cross at two points.
The surface modes from each interface can only couple around the two crossing points in
the $\omega$-$\phi$ space and this results in two resonant regions with $\xi$ being close
to $1$ [See Figure~\ref{fig4}]. The resonant regions correspond to the resonant peaks in the
spectral function shown in Figure~\ref{fig2}(c). 
Due to the mismatch of the surface modes from the two interfaces, the spectral function is
reduced and so is the HTC.   

To conclude, we have considered the situation where the near-field radiative heat transfer
between two magnetic Weyl semimetals are dominated by the nonreciprocal surface plasmon
polaritons. Due to the intrinsic nonreciprocity, the heat transfer can be effectively
controlled by a relative rotaion of parallel slabs (or twist) without surface structuring
or external field.

\begin{acknowledgments}
{\it Acknowledgments.--}
G.T. thanks Christoph Bruder for discussions and acknowledges financial support from the
Swiss National Science Foundation (SNSF) and the NCCR Quantum Science and Technology.
J.C. and L.Z. acknowledges the support from the National Natural Science Foundation of
China (Grants No. 12074230, 12047571), National Key R\&D Program of China under Grants No.
2017YFA0304203, 1331KSC, Shanxi Province 100-Plan Talent Program.
\end{acknowledgments}

\bibliography{bib_heat_radiation}{}

\clearpage

\begin{center}
  \large{\bf{Supporting Information}}
\end{center}

When the incidence plane is at azimuthal angle $\phi$ with respect to the $x$-axis, the
dielectric tensor becomes
\begin{equation}
  \bar{\bar{\epsilon}}\,{}' = {\cal R} \ \bar{\bar{\epsilon}} \ {\cal R}^T =
  \begin{bmatrix}
    \epsilon_{xx} & \epsilon_{xy} & \epsilon_{xz} \\ 
    \epsilon_{yx} & \epsilon_{yy} & \epsilon_{yz} \\ 
    \epsilon_{zx} & \epsilon_{zy} & \epsilon_{zz}
  \end{bmatrix} ,
\end{equation}
with the rotation matrix
\begin{equation}
  {\cal R} =
  \begin{bmatrix}
    \cos(\phi+\theta) & -\sin(\phi+\theta) & 0 \\
    \sin(\phi+\theta) & \cos(\phi+\theta) & 0 \\
    0 & 0 & 1
  \end{bmatrix} .
\end{equation}
For the lower WSM, we put $\theta =0$. 
We first focus on the interface between the lower Weyl semimetal (WSM) and air. The
general form of the electric and magnetic fields inside the WSM can be written as
\begin{equation}
  {\bf E} = ({\cal E}_x, {\cal E}_y, {\cal E}_z) e^{iqx -i\omega t}, \quad
  {\bf H} = ({\cal H}_x, {\cal H}_y, {\cal H}_z) e^{iqx -i\omega t}, 
\end{equation}
where the superscript $'$ in the space variables $x'$, $y'$ and $z'$ is dropped for
simplicity. From Maxwell equations, we get the differential equation
\begin{equation} \label{diff}
  d M / dz = i K M
\end{equation}
where $M=[{\cal E}_x, {\cal E}_y, \alpha{\cal H}_x, \alpha{\cal H}_y]^T$ 
with $\alpha=\sqrt{\mu_0/\epsilon_0}$ and 
\begin{widetext}
\begin{equation}
  K = 
  \begin{bmatrix}
    -q\epsilon_{zx}/\epsilon_{zz} & -q\epsilon_{zy}/\epsilon_{zz} & 0 &
    k_0- q^2/(k_0\epsilon_{zz}) \\
    0 & 0 & -k_0 & 0 \\
    k_0(-\epsilon_{yx}+\epsilon_{yz}\epsilon_{zx}/\epsilon_{zz}) &
    k_0(-\epsilon_{yy}+\epsilon_{yz}\epsilon_{zy}/\epsilon_{zz})+q^2/k_0 & 0 &
    q\epsilon_{yz}/\epsilon_{zz} \\
    k_0(\epsilon_{xx}-\epsilon_{xz}\epsilon_{zx}/\epsilon_{zz}) &
    k_0(\epsilon_{xy}-\epsilon_{xz}\epsilon_{zy}/\epsilon_{zz}) & 0 &
    -q\epsilon_{xz}/\epsilon_{zz}
  \end{bmatrix}.
\end{equation}
\end{widetext}
This differential equation has the solution 
\begin{align} 
  & [{\cal E}_x(z), {\cal E}_y(z), \alpha{\cal H}_x(z), \alpha{\cal H}_y(z)] \notag \\
  =& \sum\nolimits_{m=1}^{2}c_m[u_{1,m}, u_{2,m}, u_{3,m}, u_{4,m}] e^{ik_m z}, 
\end{align}
where $k_m$ and $u_{i,m}$ are, respectively, the eigenvalue and eigenvector of matrix $K$.
Since $K$ is a four-by-four matrix, there are four eigenvalues: two of them satisfy ${\rm
Im}(k_m)<0$ and the other two ${\rm Im}(k_m)>0$. We take $k_m$ with ${\rm Im}(k_m)<0$, of
which the subscripts are denoted as $m=1, 2$, to ensure that the electromagnetic fields
vanish at $z{\rightarrow}{-\infty}$.

The incoming electric and magnetic fields in air can be, respectively, written as
\begin{align}
  {\bf E}_{\rm in} =& \left[ e_{\rm in}^s \hat{y} +e_{\rm in}^p (\beta_0 \hat{x} -q
  \hat{z})/k_0 \right] e^{iqx +i\beta_0 z -i\omega t} , \\
  \alpha{\bf H}_{\rm in} =& \left[ e_{\rm in}^p \hat{y} -e_{\rm in}^s (\beta_0 \hat{x} -q
  \hat{z})/k_0 \right] e^{iqx +i\beta_0 z -i\omega t} ,
\end{align}
where the superscripts $s$ and $p$ are used to denote the polarization states.
The reflected fields are then expressed as
\begin{align}
  {\bf E}_{\rm re} =& \left[ e_{\rm re}^s \hat{y} -e_{\rm re}^p (\beta_0 \hat{x} +q
  \hat{z})/k_0 \right] e^{iqx -i\beta_0 z -i\omega t}, \\
  \alpha{\bf H}_{\rm re} =& \left[ e_{\rm re}^p \hat{y} +e_{\rm re}^s (\beta_0 \hat{x} +q
  \hat{z})/k_0 \right] e^{iqx -i\beta_0 z -i\omega t}.
\end{align}
At the interface of the air side with $z=0^+$, the in-plane components of the electric
and magnetic fields are
\begin{align}
  {\bf E}_{\parallel} =& [ (e_{\rm in}^s +e_{\rm re}^s) \hat{y} + (e_{\rm in}^p -e_{\rm
  re}^p) \beta_0/k_0 \hat{x} ] e^{iqx -i\omega t} , \\
  \alpha{\bf H}_{\parallel} =& [ (e_{\rm in}^p +e_{\rm re}^p) \hat{y} + (e_{\rm re}^s
  -e_{\rm in}^s) \beta_0/k_0 \hat{x} ] e^{iqx -i\omega t} .
\end{align}

For the case of the $p$-polarized incoming field, that is, $e_{\rm in}^s =0$, the
interface conditions give 
\begin{align}
  (e_{\rm in}^p -e_{\rm re}^p) \beta_0/k_0 =& c_1 u_{1,1} +c_2 u_{1,2}, \label{ep1}\\
  e_{\rm re}^s =& c_1 u_{2,1} +c_2 u_{2,2}, \label{ep2}\\
  e_{\rm re}^s \beta_0/k_0 =& c_1 u_{3,1} +c_2 u_{3,2}, \label{ep3}\\
  e_{\rm in}^p +e_{\rm re}^p =& c_1 u_{4,1} +c_2 u_{4,2}. \label{ep4}
\end{align}
From Eqs.~\eqref{ep2} and \eqref{ep3}, we have
\begin{equation}
  c_2/c_1=-(u_{2,1}\beta_0 -u_{3,1}k_0)/(u_{2,2}\beta_0 -u_{3,2}k_0).
\end{equation}
The reflection coefficient $r^{pp}=e_{\rm re}^p/e_{\rm in}^p$ can be obtained from
Eqs.~\eqref{ep1} and \eqref{ep4} as
\begin{equation}
  r^{pp} =\frac{(u_{4,1}\beta_0 -u_{1,1}k_0)+(u_{4,2}\beta_0 -u_{1,2}k_0)
  c_2/c_1}{(u_{4,1}\beta_0 +u_{1,1}k_0)+(u_{4,2}\beta_0 +u_{1,2}k_0)c_2/c_1}. 
\end{equation}
From Eqs.~\eqref{ep1} and \eqref{ep3}, we obtain $r^{sp}=e_{\rm re}^s/e_{\rm in}^p$ as
\begin{equation}
  r^{sp} =(1-r^{pp})\frac{u_{3,1}+u_{3,2}c_2/c_1}{u_{1,1}+u_{1,2}c_2/c_1}. 
\end{equation}

For the case of the $s$-polarized incoming field, that is, $e_{\rm in}^p =0$, the
interface conditions give 
\begin{align}
  -e_{\rm re}^p \beta_0/k_0 =& d_1 u_{1,1} +d_2 u_{1,2}, \label{es1}\\
  e_{\rm in}^s +e_{\rm re}^s =& d_1 u_{2,1} +d_2 u_{2,2}, \label{es2}\\
  (e_{\rm re}^s -e_{\rm in}^s) \beta_0/k_0 =& d_1 u_{3,1} +d_2 u_{3,2}, \label{es3}\\
  e_{\rm re}^p =& d_1 u_{4,1} +d_2 u_{4,2}. \label{es4}
\end{align}
Here, we have use $d_1$ and $d_2$ instead of $c_1$ and $c_2$ to distinguish from the case
of the $p$-polarized incoming field. 
From Eqs.~\eqref{es1} and \eqref{es4}, we have
\begin{equation}
  d_2/d_1 =-(u_{4,1}\beta_0 +u_{1,1}k_0)/(u_{4,2}\beta_0 +u_{1,2}k_0).
\end{equation}
The reflection coefficient $r^{ss}=e_{\rm re}^s/e_{\rm in}^s$ can be obtained from
Eqs.~\eqref{es2} and \eqref{es3} as
\begin{equation}
  r^{ss} =\frac{(u_{2,1}\beta_0 +u_{3,1}k_0)+(u_{2,2}\beta_0 +u_{3,2}k_0)
  d_2/d_1}{(u_{2,1}\beta_0 -u_{3,1}k_0)+(u_{2,2}\beta_0 -u_{3,2}k_0)d_2/d_1}. 
\end{equation}
From Eqs.~\eqref{es1} and \eqref{es3}, we obtain $r^{ps}=e_{\rm re}^p/e_{\rm in}^s$ as
\begin{equation}
  r^{ps} =(1-r^{ss})\frac{u_{1,1}+u_{1,2}d_2/d_1}{u_{3,1}+u_{3,2}d_2/d_1}. 
\end{equation}
The reflection coefficient matrix at the interface between air and WSM $n$ with $n=1,2$ is
given by
\begin{equation} \label{R}
  {\bf R}_n=
  \begin{bmatrix}
    r^{pp}_n & r^{ps}_n \\ r^{sp}_n & r^{ss}_n
  \end{bmatrix} .
\end{equation}

For the particular case of $\phi=\theta=0$, Eq.~\eqref{diff} reduces to
\begin{equation}
  \frac{d}{dz}
  \begin{bmatrix}
    {\cal E}_x \\ \alpha{\cal H}_y
  \end{bmatrix}
  = i K^{p}
  \begin{bmatrix}
    {\cal E}_x \\ \alpha{\cal H}_y
  \end{bmatrix} ,
  \quad
  \frac{d}{dz}
  \begin{bmatrix}
    {\cal E}_y \\ \alpha{\cal H}_x
  \end{bmatrix}
  = i K^{s}
  \begin{bmatrix}
    {\cal E}_y \\ \alpha{\cal H}_x
  \end{bmatrix} ,
\end{equation}
with 
\begin{equation}
  K^{p} = 
  \begin{bmatrix}
    iq\epsilon_a/\epsilon_d & k_0- q^2/(k_0\epsilon_d) \\
    k_0(\epsilon_d-\epsilon_a^2/\epsilon_d) & -iq\epsilon_a/\epsilon_d
  \end{bmatrix} 
\end{equation}
and
\begin{equation}
  K^{s} = 
  \begin{bmatrix}
    0 & -k_0 \\ -k_0\epsilon_d+q^2/k_0 & 0
  \end{bmatrix} .
\end{equation}
One can find that $\beta_1$ under $\phi=0$ satisfies $\det(\beta_1-K^{p})=0$ and the
solution to $\det(\beta_1'-K^{s})=0$ is given by $(\beta_1')^2+q^2=\epsilon_d k_0^2$ with
${\rm Im}(\beta_1')<0$.
The reflection coefficients for the $p$- and $s$-polarized modes are, respectively, given
by
\begin{equation}
  r^{pp} =\frac{\epsilon_{\rm eff}\beta_0-(\beta_1 +iq \epsilon_a /\epsilon_d)}
  {\epsilon_{\rm eff}\beta_0+(\beta_1 +iq \epsilon_a /\epsilon_d)},  \quad 
  r^{ss} =\frac{\beta_0-\beta_1'}{\beta_0+\beta_1'}, 
\end{equation}
and the other reflection coefficients vanish. The photonic transmission coefficient in
this particular case is expressed as 
\begin{equation}
  \xi(\omega,q,\phi=0) = 
  \frac{4[{\rm Im}(r^{pp})]^2 e^{-2|\beta_0|d}}{|1-(r^{pp})^2 e^{-2|\beta_0|d}|^2} +
  [r^{pp} \rightarrow r^{ss}] .
\end{equation}

\begin{figure}
\centering
\includegraphics[width=\columnwidth]{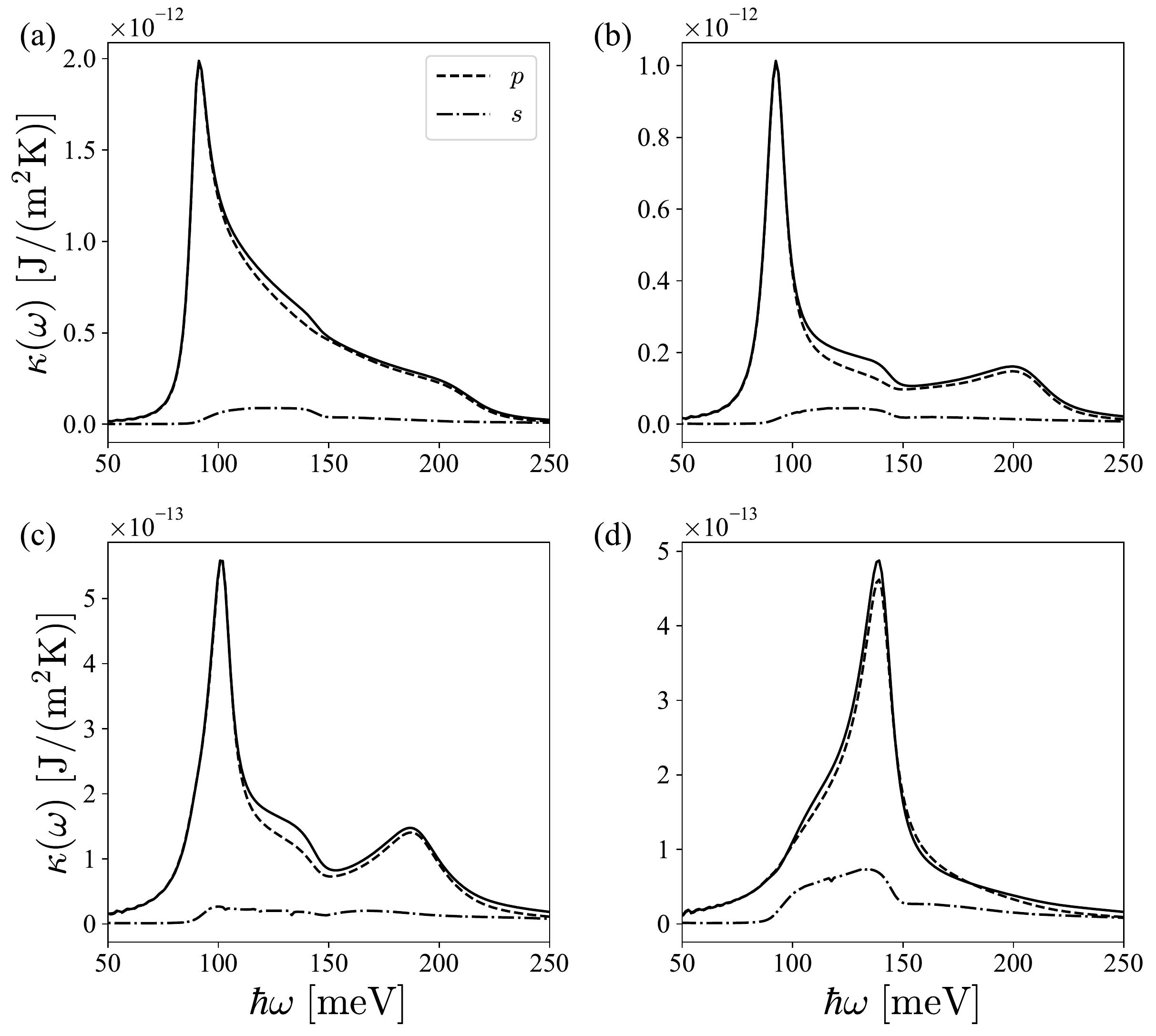} \\
\caption{Spectral function $\kappa(\omega)$ at twist angles (a) $\theta=0$, (b)
$\theta=\pi/4$, (c) $\theta=\pi/2$, and (d) $\theta=\pi$ with $d=100\,$nm. The solid lines
are obtained by considering the contributions from all the elements in the reflection
coefficient matrix ${\bf R}_n$. The dashed (dash-dotted) lines give the contribution from
the $p$-polarized ($s$-polarized) mode by setting $r_n^{ps}=r_n^{sp}=r_n^{ss}=0$
($r_n^{ps}=r_n^{sp}=r_n^{pp}=0$) in ${\bf R}_n$. It can be infered from the lines that the
$p$-polarized mode dominates the near-field radiative heat transfer. } 
  \label{fig_SM}
\end{figure}

\end{document}